\documentclass[12pt]{iopart}
\usepackage{iopams}
\usepackage{rotating}
	\usepackage{dblfloatfix}
\usepackage{graphicx}
\usepackage{epsfig}
\usepackage[normalem]{ulem}  

\begin{document}
\title{Properties of Massive Rotating Protoneutron Stars with Hyperons: Structure and Universality}
\author{Smruti Smita Lenka$^{\rm (a)}$, Prasanta Char$^{\rm (b)}$, Sarmistha Banik$^{\rm (a)}$}
\address{$^{\rm (a)}$BITS Pilani, Hyderabad Campus, Shameerpet Mandal, Hyderabad 500078, India}
\address{$^{\rm (b)}$INFN Sezione di Ferrara, Via Saragat 1, I-44100 Ferrara, Italy}

\begin{abstract}
In this work,  we study the properties and structure of a massive and rapidly rotating protoneutron star (PNS)  with hyperon content. We follow several stages of quasi-stationary evolution in an approximate way at four discrete steps. We use a density-dependent (DD) relativistic mean field theory (RMF) model and calculate different quantities such as mass, equatorial radius, moment of inertia, and quadrupole moment to get different rotating configurations upto the mass-shedding limit. We study the effect of the appearance of $\Lambda$, the lightest of all hyperons, on each of the evolutionary stages of the PNS. We also check its sensitivity to the inclusion of $\phi$ vector meson as a mediator of $\Lambda-\Lambda$ interaction in detail. Finally, we investigate the universal relations between moment of inertia and compactness in the context of a hot and young compact object.
\end{abstract}
\maketitle
\submitto{\jpg}

\section{Introduction} 
When a massive star ($M \gtrsim 8M_{solar} $) reaches the end of its life, its core collapses. This leads to a supernova explosion leaving  its center as a dense compact remnant, called a PNS. Initially, the PNS is very hot, lepton-rich and rapidly rotating. It deleptonizes releasing the trapped neutrinos. In the process, the neutrinos heat up the PNS while decreasing the net lepton fraction. After that, the PNS enters a steady cooling phase and becomes a cold, catalyzed neutron star (NS) \cite{proto,PrakashPR97,Burr86}.  The PNS is believed to have four prominent stages of evolution as it ends up as a cold catalyzed object \cite{PrakashPR97,Burr86}. These are governed strongly by the nature of matter at very high density as well as the neutrino reaction rates and diffusion timescales. There have been many detailed studies on the global properties of the PNS with different microphysical inputs \cite{pons3,pons2,Keil95}. The  simplified models of post-bounce neutrino emission \cite{Burr86,Keil95} have  advanced  to full solutions of the Boltzmann equation \cite{Fisher}, with more realistic microphysics \cite{Hud} in the last few years.  Many of these studies of PNS cooling have used the equilibrium flux-limited diffusion (EFLD) approximation, while the variable Eddington factor method has been used for neutrino transport \cite{Robert}.

At the high density core, the Fermi energy of nucleons become sufficiently large; according to Pauli principle, the formation of hyperon becomes energetically favorable. The $\Lambda$ hyperons, being the lightest among other massive baryons are eventually the first to populate the core as the nucleon chemical potential outweighs their in-medium mass \cite{Gle}. The in-medium mass of hyperon is determined from experimental inputs such as the nature of the nuclear-hyperon interaction. Several studies have been done to investigate the effect of hyperons on the structure of PNS \cite{Bednarek2006, Mu2017}. But, they have used mostly older equation of state (EOS) parametrizations which are not consistent with the recent observations or the latest experimental data. Therefore, it is our objective to revisit the problem with more suitable microphysical inputs and find out the sensitivity of these results with respect to the parameters of the theory. In this work, we will mainly focus on the dense nuclear matter containing hyperons and study its effect on global properties of massive PNS at four different stages of evolution, starting from neutrino-trapped, isoentropic PNS to cold catalyzed $\beta$-equilibrated NS. Keeping that in mind, we have employed a realistic EOS including hyperons for this work. Incidentally, we have also constructed new EOSs  for the neutrino-trapped, isentropic matter and use them to study the initial stages of PNS evolution.  As a matter of fact, only the electron-type neutrinos become trapped during the core collapse. Therefore, we have considered only the electron-type neutrinos in our study \cite{PrakashPR97}.

The EOSs are constructed using a relativistic hadron field theoretical model with density dependent couplings in the mean field approximation. We use the DD2 parameter set for the couplings \cite{typel05,Typel2010}. It satisfies the constraints on nuclear symmetry energy and its slope parameter as well as the incompressibility from the nuclear physics experiments \cite{dd2_14}. Furthermore, the astrophysical observations also give us important clues about the EOS of dense matter. The recent detection of gravitational wave signal from the binary neutron star merger event GW170817 provides the latest information in this regard \cite{Abbott2017}. Besides, the $\sim 2$ $M_{solar}$ mass measurement of PSRs J1614-2230 and J0348+0432 has severely constrained the parameter space for the NS EOS \cite{demo,anton,fons}. Now, the cold NS mass from hyperonic DD2 model is $2.1$ $M_{solar}$. Thus, it satisfies the observational limit  on the maximum mass of the NS. It is also observed that it satisfies the tidal deformability bounds found from GW170817 \cite{sb_2017}. 

Another objective of this work is to study the universal relations recently discovered among various global quantities of compact objects such as normalized moment of inertia ($\bar{I}:=I/M^3$, $I/MR^2$) and
stellar compactness ($\mathcal{C}:=M/R$)  \cite{yagi2013,lattimer05,breu2016}.
In the context of the PNS evolution, $\bar{I}-\bar{Q}$ relation has been studied by Martinon et al. \cite{Marti2014} for nucleon-only EOS. They have found that the universality is broken during the initial stages after the core bounce, but it is satisfied at later stages. Recently, Marques et al. \cite{Marq17} also investigated the problem for rapidly rotating hot stars with hyperonic EOS. They found that the relation does not change in the presence of hyperons but deviates for high entropy. However, the $\bar{I}$ vs $\mathcal{C}$ relations has not been studied for the PNS till now.  Previously, Breu and Rezzolla  studied a large set of cold nucleonic EOSs with different stiffness and showed that universality relation holds for normalized moment of inertia \cite{breu2016}. We studied these relations in the presence of exotic components like hyperons and antikaon condensates for cold NS \cite{ijmpd}. In this work, we will examine the $\bar{I}$ vs $\mathcal{C}$  relations for PNS using the fitting factors provided by Breu and Rezzolla \cite{breu2016}. 

The paper is organized in the following way. In section \ref{Evolution}, we describe the PNS evolution scenarios, followed by section \ref{EOS-PNS} explaining the EOSs of dense matter. In section \ref{lorene}, we discuss the structure of rotating relativistic star. Finally, in section 
\ref{result}, we discuss our results and conclude with a summary in section \ref{sum}.

\section{Stages of PNS Evolution}\label{Evolution}
Within  a  few  milliseconds  after  the  core  bounce,  the  hydrostatic
equilibrium  is  reached  in  the  lepton-rich core and the star enters the Kelvin-Helmholtz cooling phase which lasts for about 10s \cite{PrakashPR97, Burr86, Keil95}.  In this phase, the evolution is completely determined by the neutrino diffusion. This is the formation phase of the star. The central temperature and entropy keep increasing despite neutrino loss.  The whole star starts to cool down as a whole and the total entropy decreases significantly at 20s. It fully deleptonizes at about 30s \cite{proto} and  the neutrinoless $\beta$-equilibrium is reached. Ultimately, after one minute the temperature drops down to $\sim 1$ MeV. 
We follow a well established evolutionary scenario \cite{PrakashPR97, Burr86} 
that suggests that the PNS undergoes roughly four stages of evolution towards becoming a stable, cold catalyzed compact object. This picture has been used by many people to study PNS \cite{Bednarek2006, Mu2017, Veronica}.
\begin{enumerate}
\item Just after its birth, the PNS initially has trapped neutrinos and the lepton fraction is $Y_L = 0.4$. The core of the PNS has an entropy per baryon $s_B = 1$ surrounded by a high entropy, neutrino trapped outer layer, which deleptonizes faster than the core.
\item While the outer layer is being deleptonized, the central object is still neutrino trapped with $Y_L = 0.4$ and the neutrino diffusion heats it up to $s_B = 2$.
\item After complete deleptonization, the core becomes neutrino-free and attains high entropy ($Y_{\nu} = 0, s_B = 2$).
\item Finally, the star settles as a cold stable NS in beta equilibrium.
\end{enumerate}
In the results section, we will denote these four stages with I to IV. Our aim is to study the properties and structure of the star at each of these discrete stages. This way the quasi-stationary PNS evolution has been approximated  to reproduce qualitatively the different evolutionary stages. However, this approximate scheme has its own limitation. It does not represent a complete picture of structural evolution of the PNS during the cooling phase. We consider G = c = $k_B$ = 1 throughout this work, G, c and $k_B$ being the universal gravitational constant, speed of light and Boltzmann constant respectively.

\section{Equation of State of PNS matter}\label{EOS-PNS}
 
 In this section, we briefly discuss some of the important features of the formalism, we follow.  Our PNS EOSs are constructed in a unified way by smoothly matching the high density and low density parts following Banik et al. \cite{Banik2014}.  For the low-density part, nuclear statistical model of Hempel and Schaffner-Bielich \cite{Hempel2010}  is used to treat the heavy and light nuclei as well as the interacting nucleons, including the excluded volume effects and other in-medium effects. The high dense matter, which is made up of neutrons, protons,  electrons and muons was developed by Hempel and Schaffner-Bielich and is denoted by HS(DD2) \cite{Hempel2010}. It  is described by DDRMF model, where the interaction among baryons is mediated by $\sigma$, $\omega$ and $\rho$ mesons. The high density core is expected to consist of hyperons as well. There is an additional vector meson ($\phi$) which accounts for the hyperon-hyperon interaction. 
The EOS with $\Lambda$-hyperons is denoted by BHB$\Lambda$. In the presence of hyperon-hyperon interaction via $\phi$ mesons, the EOS is represented by BHB$\Lambda \phi$ \cite{Banik2014}. For the lepton-trapped PNS, we have considered only electron-type neutrinos, in addition to the hadrons and electrons. As mentioned in Section II,
in the neutrino-trapped matter, the electron lepton number $Y_{Le}= Y_e+ Y_{\nu_e} =0.4$ \cite{PrakashPR97}. Also, as no muons are present
when the neutrinos are being trapped, the constraint $Y_{L\mu}= Y_{\mu} + Y_{\nu_{\mu}} =0$ can be imposed. Hence we fix the lepton fraction $Y_L=0.4$ in our calculation.
A finite contribution of neutrinos make considerable difference to the EOS. 

The nucleon-meson couplings in the present model are density dependent. The functional form for this dependence and the corresponding parameter set (DD2) is taken from Typel et al. \cite{Typel2010}. The saturation properties of the DD2 parameter set are in good agreement with the nuclear physics experiments \cite{Lattimer2013}. The hyperon-vector meson couplings are calculated from the SU(6) symmetry relations \cite{Schaffner1996} and hyperon-scalar meson coupling is determined from the hypernuclei data which gives the hypernuclei potential depth in the normal nuclear matter \cite{Chrien1989}.

We have considered a
potential of depth $U^{(N)}_{\Lambda}(n_0) \sim −28$ MeV, as it reproduces the
bulk of $\Lambda$-hypernuclei binding energies \cite{Mill88}.
The potential $\Sigma$ hyperons feel in nuclear matter is quite uncertain. The  analyses of ($\pi^-$, $K^+$) reactions off nuclei suggest a moderate repulsive potential, whereas  $\sigma^-$-atomic data indicates an attractive potential at the surface to a repulsive one inside the nucleus. The experimental uncertainties can vary the potential depth of $\Xi$-hyperons in symmetric nuclear matter from -18 MeV to 0 MeV \cite{Tolos}.
It was found that the uncertainties in hyperon-nuclear interaction affect the appearance of the different hyperons strongly. However, they do not cause much difference to the maximum masses \cite{Tolos}. Hence 
heavier hyperons such as $\Sigma$ and $\Xi$ are not considered in our studies \cite{Banik2014}. 
The experimental studies of HICs \cite{Mori15}
and the theoretical studies derived from chiral effective forces \cite{Hai17}
are expected to narrow down these uncertainties of  hyperon-nucleon interactions in dense matter.\\ 

\section{Rotating Star Structure}\label{lorene}
We compute the structure of rotating stars using  LORENE \cite{lorene}
code which assumes a stationary, axisymmetric spacetime. The line element is given by,
\begin{equation}
g_{\alpha \beta}dx^{\alpha} dx^{\beta}= -N^2 dt^2 + A^2 \left( dr^2 + r^2 d \theta^2\right) + B^2 r^2 sin^2\theta {\left(d \phi - \omega dt\right) }^2,
	\label{eq_metric}
\end{equation}
where, N, A, B and $\omega$ are functions of (r, $\theta$). The energy-momentum tensor for a perfect fluid is related to energy density $\epsilon$, pressure P and four-velocity $u^{\mu}$ by,
$T^{\mu\nu}=(\epsilon+ P)u^{\mu}u^{\nu} + P g_{\mu\nu}$.
The equation for stationary motion is given by \cite{Gous97}, 
\begin{equation}
\partial_i\left(H+ln\frac{N}{\Gamma}\right) = T e^{-H}\partial_i s_B,
\end{equation}
where, $s_B$ is the entropy per baryon, T is the temperature, $H = ln \left(\frac{\epsilon + P}{n_B M_B}\right)$ is the log-enthalpy, $\Gamma = (1 - U^2)^{-1/2}$ is the Lorentz factor, U is the fluid velocity. The quantities U, $\epsilon$ and P are measured by locally non-rotating observer. LORENE is primarily formulated for cold EOS, or a barotropic EOS \cite{Gous97}. Our EOSs for PNS are temperature dependent as we have considered isoentropic profile for the stars.
This homoentropic flow makes the EOS barotropic, which enables us to use the LORENE formalism to study the PNS stages. 
                                
The gravitational mass, angular momentum and quadrupole moment are given respectively as \cite{salgado,Pappas,Friedman},
\begin{equation}
 M = \frac{1}{4 \pi} \int \sigma_{ln N} ~ r^2 \sin^2 \theta dr d\theta d\phi ~
\end{equation}
\begin{equation}
J = \int A^2 B^2 (E + P) U r^3 \sin^2 \theta dr d\theta d\phi ~.
\end{equation}
\begin{equation}
 Q = - M_2 - \frac{4}{3}\left(b+ \frac{1}{4}\right) M^3~, 
\end{equation}
  where,
\begin{equation}
M_2 = - \frac{3}{8\pi} \int \sigma_{ln N} \left(\cos^2 \theta - \frac{1}{3} \right) r^4 \sin^2 \theta dr d\theta d\phi ~
\end{equation}
Here, $\sigma_{ln N}$ is the source term in the expression for $ln N$ (given by the RHS of Eq. $3.19$ of \cite{BGSM}); $E = \Gamma^2(\epsilon$ + P) - P.  The quantity $b$ is defined by \cite{Friedman},
\begin{equation}
 b =- \frac{8}{\pi M^2} \int P r \sin \theta e^{\nu} dV
\end{equation}
dV being the volume element.
Then, the moment of inertia of the rotating star is defined as, $I=J/\Omega$, where $\Omega$ is the stellar angular velocity.

\begin{table}
\caption{\label{tab1}Gravitational Mass-Radius for static stars.} 
\centering
\par
\begin{tabular}{|c | c | c | c |}
  \hline

 EOS & Evolution Stages  & $M$ ($M_{solar}$)& $R$ (km)  \\

 \hline
 HS(DD2)  & $s_B$=1, $Y_L$=0.4   & 2.375 & 12.610 \\
 & $s_B$=2, $Y_L$=0.4   & 2.390 & 13.181 \\
 & $s_B$=2, $Y_{\nu}$=0   & 2.437 & 12.889 \\
 & T= 0   & 2.423 & 11.869 \\
\hline
BHB$\Lambda$ & $s_B$=1, $Y_L$=0.4    &  2.108  &  13.044 \\
 & $s_B$=2, $Y_L$=0.4    &  2.150  &  13.888 \\
 & $s_B$=2, $Y_{\nu}$=0    &  2.018  &  12.713 \\
 & T= 0   &  1.955  &  11.737   \\
\hline
BHB$\Lambda$$\phi$ & $s_B$=1, $Y_L$=0.4     &  2.172 &  12.779  \\
& $s_B$=2, $Y_L$=0.4     &  2.202  &  13.588  \\
& $s_B$=2, $Y_{\nu}$=0     &  2.126  &  12.488  \\
& T= 0   &  2.1  &  11.608  \\

    \hline
  \end{tabular}
\end{table}

\section{Results and Discussions}\label{result}
We study the properties of a massive PNS using DD2 EOS at four different configurations (I - IV) relevant to the PNS evolution mentioned in section II. Our aim is to study how the emergence of $\Lambda$ hyperons and their interaction affect the PNS stages. We use three types of EOSs : i) HS(DD2), ii) BHB$\Lambda$, and iii) BHB$\Lambda\phi$ for our calculations, as discussed in Section III. 

\begin{figure*}[b]
\includegraphics[width=\textwidth, height=10cm]{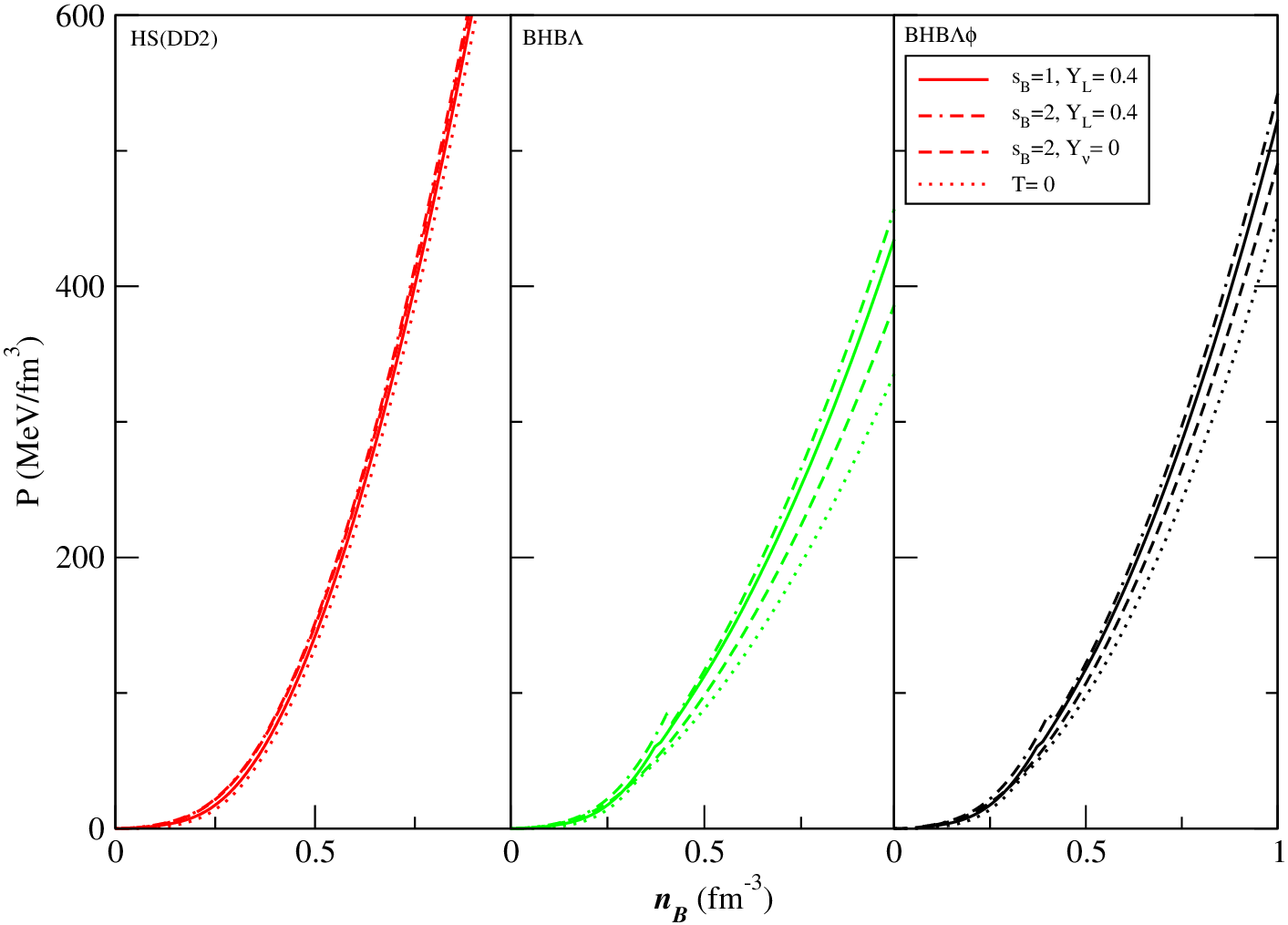}
\caption{Pressure versus number density is plotted for discrete evolutionary stages of the compact star. The three panels from left to right are for HS(DD2), BHB{$\Lambda$} and BHB{$\Lambda\phi$} EOSs respectively.}
\label{Fig1}
\end{figure*}

  In Fig.~\ref{Fig1}, pressure (P) is plotted against number density ($n_{\rm B}$). The left panel is for nucleons-only HS(DD2) EOS, the middle and the right panels are for EOSs with hyperons, namely BHB$\Lambda$ and BHB$\Lambda \phi$. The presence of hyperons softens the EOS. The solid line represents the stage I of PNS evolution i.e. lepton trapped with s$_B=1$. As the entropy per baryon increases to s$_B=2$, the EOS (dot-dashed) gets stiffer for all the three variants. Once the neutrinos leave the system, the EOS is softened again (dashed line). The EOS (dotted line) for the cold catalyzed matter is the softest among all. This difference in stiffness is pronounced largest for the BHB$\Lambda$ EOS, lesser for BHB$\Lambda\phi$, and least for HS(DD2). 
Neutrinos are produced in large number in the presence of $\Lambda$ hyperons. This was observed earlier in Ref. \cite{PrakashPR97}.
Consequently, as the PNS passes from stage II to III, the loss of lepton pressure softens the EOS considerably for BHB$\Lambda$ and BHB$\Lambda\phi$. If we compare the EOS for any particular stage, it becomes evident that the HS(DD2) is always the stiffest among the three, followed by BHB$\Lambda\phi$ and BHB$\Lambda$ respectively. The difference in stiffness among the four evolutionary stages is most pronounced for the softest EOS.

\begin{figure}
\includegraphics[width=\textwidth]{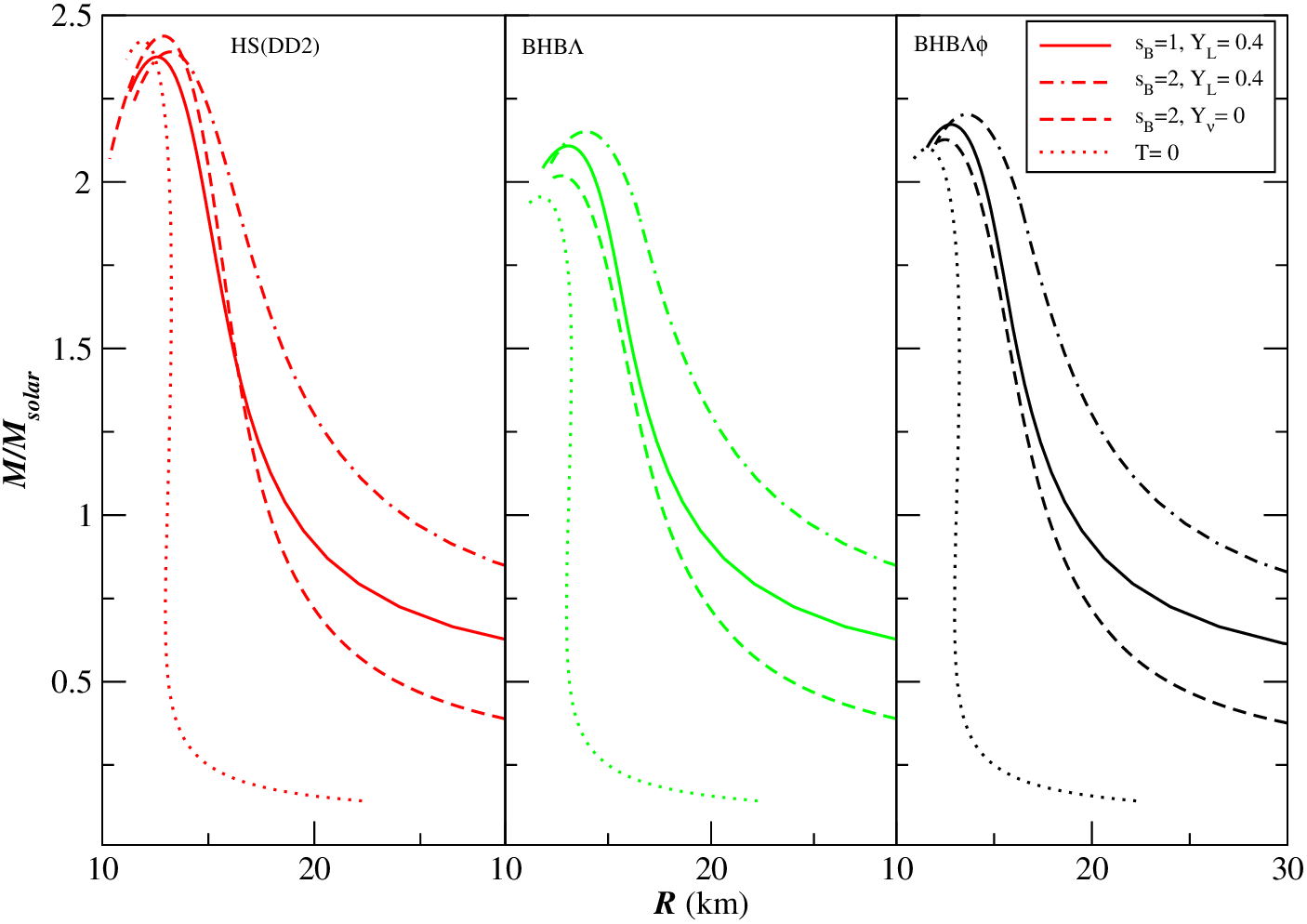}
\caption{Gravitational mass versus radius for the corresponding EOSs of Fig.~\ref{Fig1}.}
\label{Fig2}
\end{figure}

Gravitational mass ($M$) versus radius ($R$) for the corresponding EOSs of Fig.~\ref{Fig1} are plotted in Fig.~\ref{Fig2}. Here also, we follow the same line style as Fig.~\ref{Fig1}. We find that the stiffer EOS yields a higher maximum mass, as expected. However, the difference in their corresponding radii is not so prominent. The maximum gravitational mass and their corresponding radii for static configuration are given in Table \ref{tab1}. We also notice that the hotter stars with comparatively smaller masses have larger radii than their cold counterparts. This can be attributed to the higher thermal pressure of the hot stars. If we compare the $s_B$= 2 cases, a rise of 6.5(3.5)\% in maximum mass and 9.2(8.8)\% in the corresponding radius are noticed for BHB$\Lambda$( BHB$\Lambda\phi$) EoS from neutrino-free to neutrino-trapped matter.  Finally, our results corroborate the general notion that in nucleon-only system neutrino-trapping reduces the maximum mass of the star, which reverses in the presence of exotic matter such as hyperons, quarks and antikaons \cite{PrakashPR97, meta}.

\begin{figure}
\includegraphics[width=\textwidth]{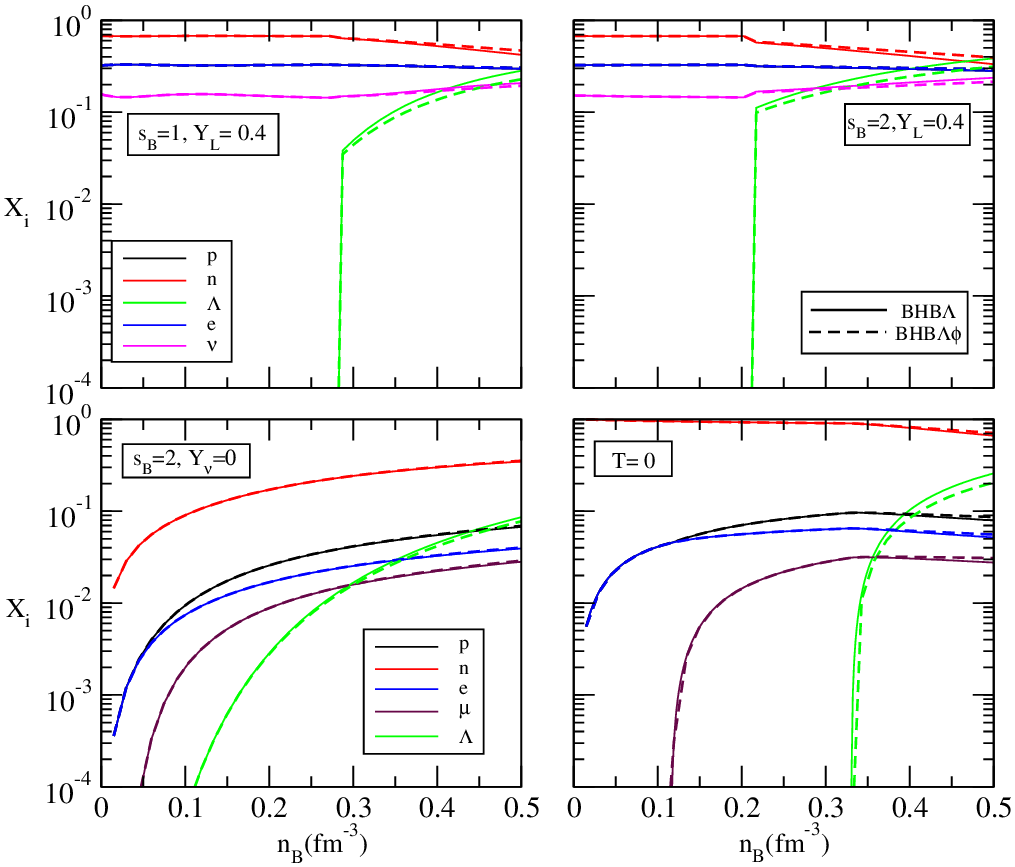}
\caption{Particle fraction variation with number density for BHB{$\Lambda$} and BHB{$\Lambda\phi$} EOSs. Upper panels are for stage-I and II whereas lower panels are for stage-III and IV.}
\label{Figfrac}
\end{figure}

In Fig.~\ref{Figfrac} we plot the particle fractions of the BHB$\Lambda$ (solid lines) and BHB$\Lambda \phi$ (dashed lines) models as functions of the baryonic density at four evolutionary stages. The upper panels are for the lepton-trapped cases (stages I and II), where we have protons, neutrons, electrons and neutrinos in addition to the $\Lambda$-hyperons. As the $\Lambda$-hyperons are produced at cost of the neutrons, the neutron fraction is noticed to drop a little at their onset. Hyperon-hyperon interaction reduces lambda fraction compared with the case without hyperon-hyperon interaction. The chemical equilibrium condition $\mu_n-\mu_p=\mu_e-\mu_{\nu}$ for the lepton-trapped PNS consequently requires the neutrino chemical potential to increase during the process. On the other hand, the charged particle fraction slightly goes down in order to maintain the constant lepton fraction of $Y_L=0.4$.  For the neutrino-free $\beta$-equilibrated matter (stages III and IV), we have muons also. At $s_B=2$, hyperons appear at lower density than zero-temperature case. For stage IV, in the presence of neutral $\Lambda$-hyperons, the drop in neutron chemical equilibrium demands a dip in both the lepton populations. To maintain charge neutrality the proton fraction also goes down. The composition of matter has a direct consequence on the stiffness of the EoS. The electron fraction is high in neutrino trapped matter, which makes the matter more proton-rich, leaving the EoS for neutrino-trapped matter stiffer than the one for neutrino-free matter. Also, the threshold of hyperons is changed in neutrino-trapped matter significantly.  In the neutrino-trapped matter, the hyperons appear at higher baryon density compared to neutrino-free matter. The exact threshold values are given in Table 2. If we compare the values for a particular entropy per baryon, say $s_B$= 2, we notice a delay of  ~90\% in threshold baryon density  of  $\Lambda$ hyperons.

\begin{figure}
\includegraphics[width=\textwidth]{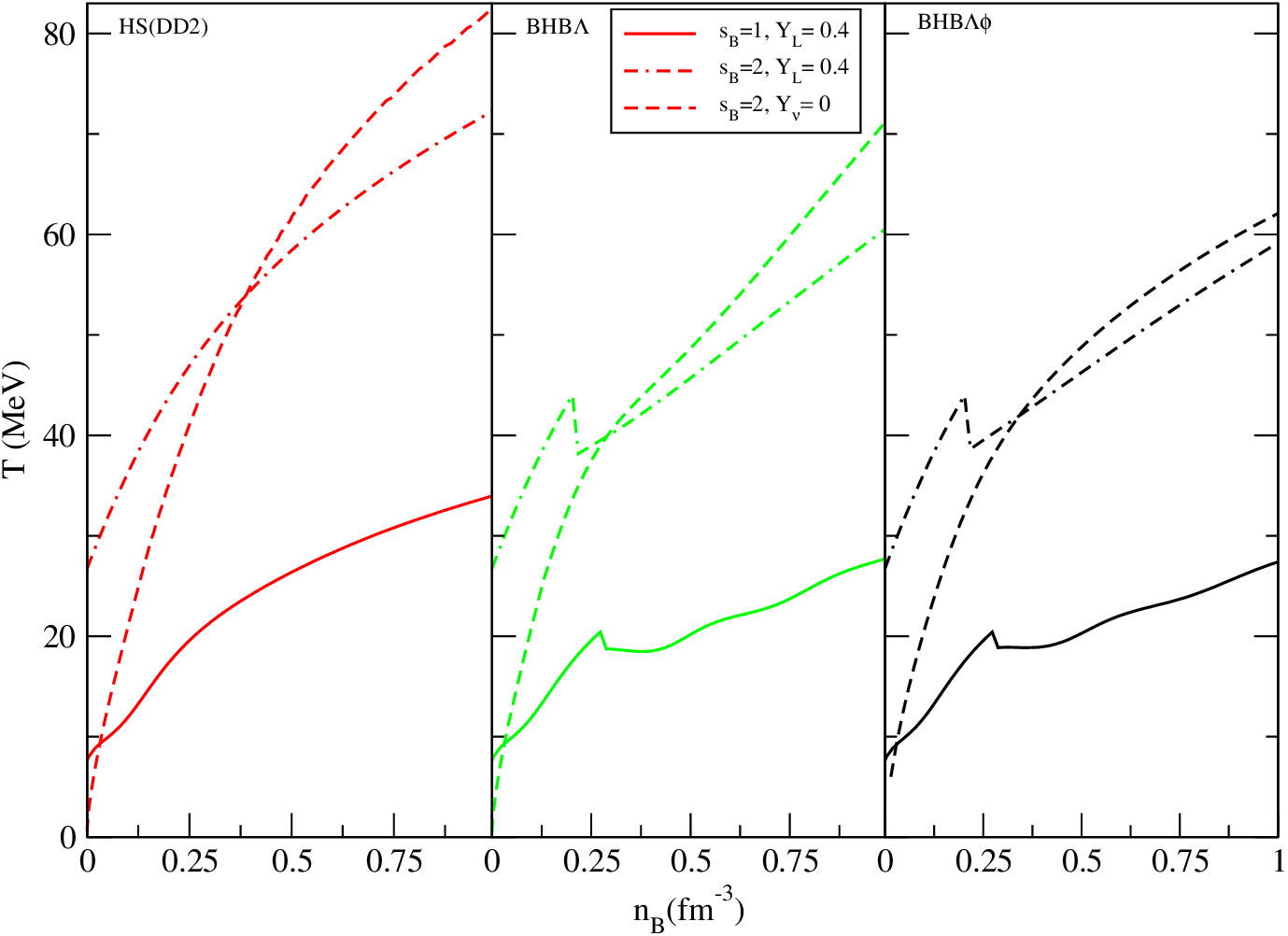}
\caption{Temperature profile of the compact star as it evolves from $s_B=1$, $Y_L=0.4$ to neutrinoless $\beta$-equilibrated neutron star of $s_B=2$, for different compositions of matter.}
\label{Fig3}
\end{figure}

The variation of temperature  as a function of baryon density is plotted in Fig.~\ref{Fig3} for the first three stages of PNS evolution. In each case, the  star is hot in the central region; the temperature falls off monotonically towards the surface. In individual panels of Fig. \ref{Fig3}, we follow the change in temperature for a particular EOS as the PNS evolves. The entropy per baryon of the PNS increases from s$_B=1$ to s$_B=2$ leading to a significant increase in the temperature as well as the radius. Inclusion of hyperons lowers the temperature of the central region of the PNS. For example, at $n_B$ $\sim$ $1~  fm^{-3}$, for HS(DD2) EOS, the temperature increases from $\sim  33$ to $\sim 68$ MeV from stage I to II, whereas in the presence of $\Lambda$ hyperons, this rise is from $\sim 26$ to $\sim 54$ MeV. The kinks in the temperature mark the appearance of $\Lambda$ hyperons. The corresponding threshold densities are listed in Table \ref{tab2}. This characteristic is quite expected. The additional degree of freedom, $\Lambda$, comes into the system with very low momentum. Hence, the temperature, which is the average kinetic energy per degree of freedom, will drop. The star size decreases when the neutrinos leave the system i.e. from Stage II to III, hence the temperature increases for all the EOS due to compression. The neutrinos carry away most of the energy. Consequently the star becomes cold at stage IV.

\begin{table}
\caption{\label{tab2}Threshold densities of $\Lambda$ hyperons (in $fm^{-3}$)}
\centering
\par
\begin{tabular}{|c|c|c | c | }
  \hline
 Evolution Stages&BHB$\Lambda$&BHB$\Lambda \phi$\\
  
  \hline
$s_B$=1, $Y_L$=0.4   &0.285&0.287\\
$s_B$=2, $Y_L$=0.4   &0.217&0.217\\
$s_B$=2, $Y_{\nu}$=0 &0.114&0.114\\
T= 0                 &0.325&0.329\\
   \hline
   
  \end{tabular}
\end{table}

\begin{figure}
\includegraphics[width=\textwidth]{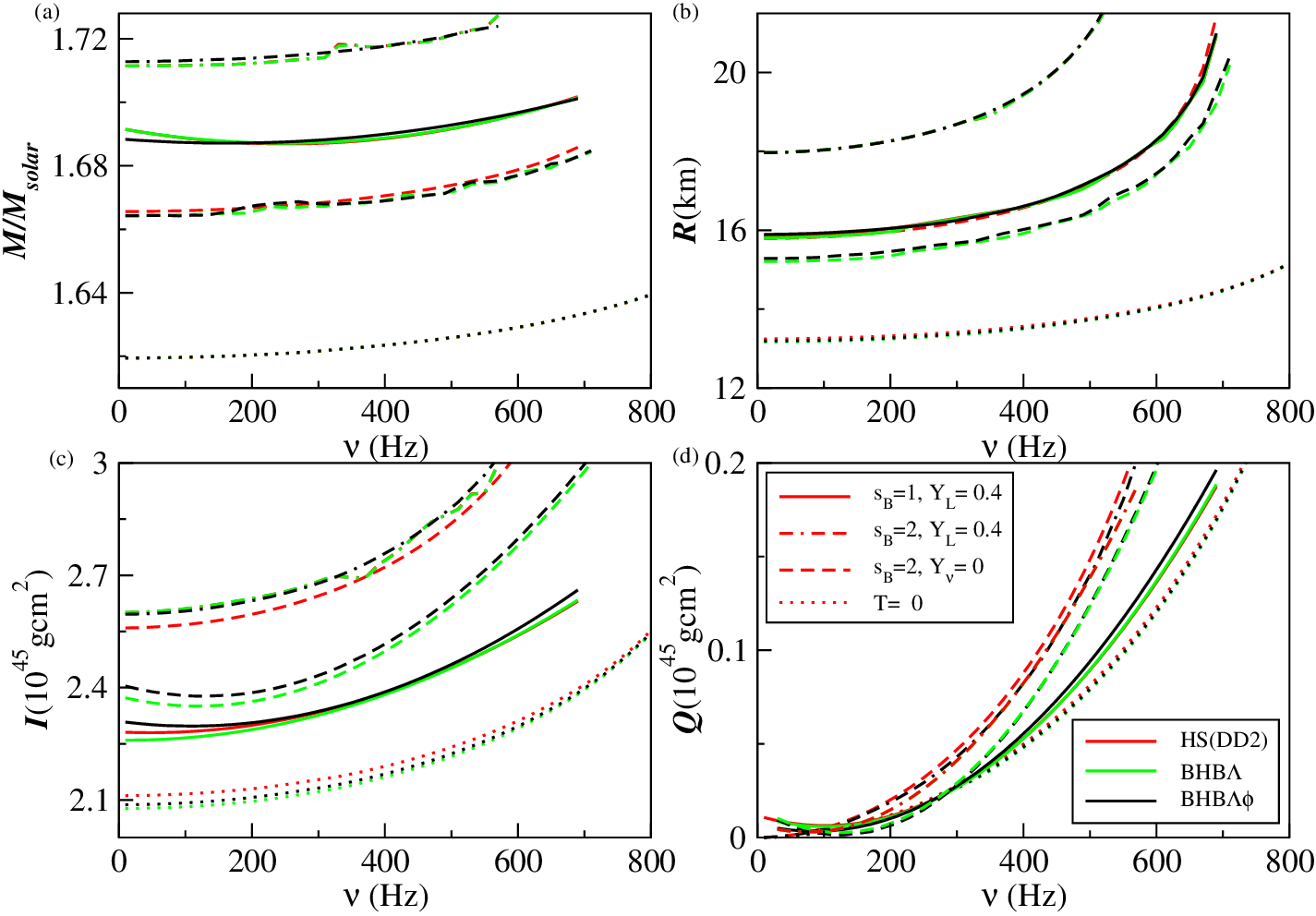}
\caption{Gravitational mass, radius, moment of Inertia and quadrupole moment of a star evolving according to Sec. \ref{Evolution} are plotted in (a), (b), (c) and (d) respectively as function of rotation frequency. All plots refer to a star with fixed baryonic mass $M_B$=1.8$M_{solar}$.}
\label{Fig4}
\end{figure}

Next, we study the effect of rotation on various stellar properties. Soon after its birth in a gravitational collapse, neutrino-trapped PNS rotates differentially. As it settles into $\beta$-equilibrium, viscosity dampens the non-uniformity in rotation. We consider an idealised scenario of uniform and rigid rotation about an axisymmetric axis which represents an approximation to the actual rotational state of a hot NS \cite{Gous97}.
For a representative purpose, let us consider a star with baryonic mass $M_B=1.8$ $M_{solar}$ and follow  its different properties such as $M$, $R$, moment of inertia ($I$) and quadrupole moment ($Q$). In Fig.~\ref{Fig4}, these quantities are plotted as a function of rotational frequency  for stages I-IV, up to the corresponding Kepler limit. All the quantities are observed to increase monotonically with rotation. This nature was noted in earlier work also \cite{Marti2014}. We have checked the stability of the stellar configurations as well. For the rigidly rotating stars with a constant total entropy $S=s_B  M_B$, the stability condition is given by  $\left(\frac{\partial J}{\partial n_B}\right)_{M_B,S} < 0$ . The angular momentum ($J$) for all the cases considered here satisfies this condition upon varying the central baryon number density until they become unstable at extremal points \cite{Marq17}. The Kepler limit for a cold-catalysed star with baryon mass of $M_B=1.8$ $M_{solar}$  is 935 Hz, while for the newly born star it is much less i.e. $\sim 703$ Hz. Initially the PNS is rotating at a lower frequency, only to reach a higher rotation rate as it contracts and cools down to a cold catalyzed $\beta$-equilibrated NS. This can be attributed to the conservation of angular momentum ($J=I\Omega$), which restricts the PNS with large $I$ to rotate slowly. Interestingly, as the PNS attains a higher entropy in the immediate step of evolution with $s_B=2$ and $Y_L=0.4$, it has to slow its rotation rate as this star can only withstand a mass-shedding frequency limit of $\sim577$ Hz. However, it can increase its rotation rate in the later stages of evolution as is evident from Fig.~\ref{Fig4}. The Kepler limit for all the EOSs in different evolution stages are given in Table \ref{tab3}. The gravitational mass remains almost independent of EOS in all the evolutionary stages. We notice an EOS-dependent spread in radii of the stars. This spread is widest for the intermediate stages, but not so explicit for initial and final stages. This is also reflected in the plots for moment of inertia and quadrupole moment versus frequency. Their values differ with their constituents for the higher entropy stages due to the differences in the radii. Both $I$ and $Q$ are less for the cold stars compared to the earlier stages. 

We tabulate the global structural properties of both non-rotating and maximally rotating PNS for a fixed baryonic mass $M_B=1.8$ $M_{solar}$ in Table \ref{tab3}. We can see that central density decreases from stages I to III for HS(DD2) EOS and then increases after the star attains $\beta$-equilibrium for both static and Keplerian scenarios. However, for the BHB$\Lambda$ and BHB$\Lambda\phi$ EOSs, the central density starts increasing after deleptonization i.e. from stage III onwards. This can be explained in the following way. As the neutrinos carry away most of the binding energy of the system, after deleptonization, the star contracts and as a result the central density increases for both non rotating and rotating stars. The angular  frequency increases for the rotating star thereby increasing the Kepler frequency. Emergence of hyperons in these two EOSs leads to greater neutrino emission resulting in an early contraction. 
In all stages, the central density for the Keplerian star is always lesser than their static counterparts for all the EOSs considered.

The gravitational mass and radius increase from stage I to stage II due to increase in thermal pressure, but decrease in the subsequent stages. This can be attributed to the loss of neutrino pressure from stage II to III and drop in thermal pressure from stage III to IV. This behavior is observed for both non-rotating and Keplerian cases and also for all EOSs. The values of $M$ and $R$ in the Table \ref{tab3} agree with the results of Fig. \ref{Fig4}. The angular momentum corresponding the Kepler limit 
changes the same way as the Kepler frequency already discussed in connection with Fig. \ref{Fig4}. Finally, we also see the value of the ratio of rotational energy to gravitational energy (${\cal T}/|W|$) steadily increasing from 0.051 to 0.115 during the evolutionary stages, with the exception in stage II. However, these values are  rather insensitive to the choice of EOS. This shows the increase in the rotational kinetic energy which leads to rise in ellipticity as is evident from  Table \ref{tab3}. Interestingly, the change in ellipticity is also independent of EOS. Similarly we have calculated the same set of quantities for a star with fixed baryonic mass $M_B= 2.2$ $M_{solar}$  and  they are listed in Table \ref{tab4}. The results are qualitatively similar to those of baryonic mass $M_B=1.8$ $M_{solar}$. These results are consistent with those found in earlier studies \cite{Villain04}. 
In Table I and II of Ref. \cite{Villain04}, the changes in central density, $M$ and $R$ with increasing time steps are listed, while we report our results for different aforementioned stages in Table III and IV. However, we use isentropic profiles unlike theirs.

\begin{figure}
\includegraphics[width=\textwidth]{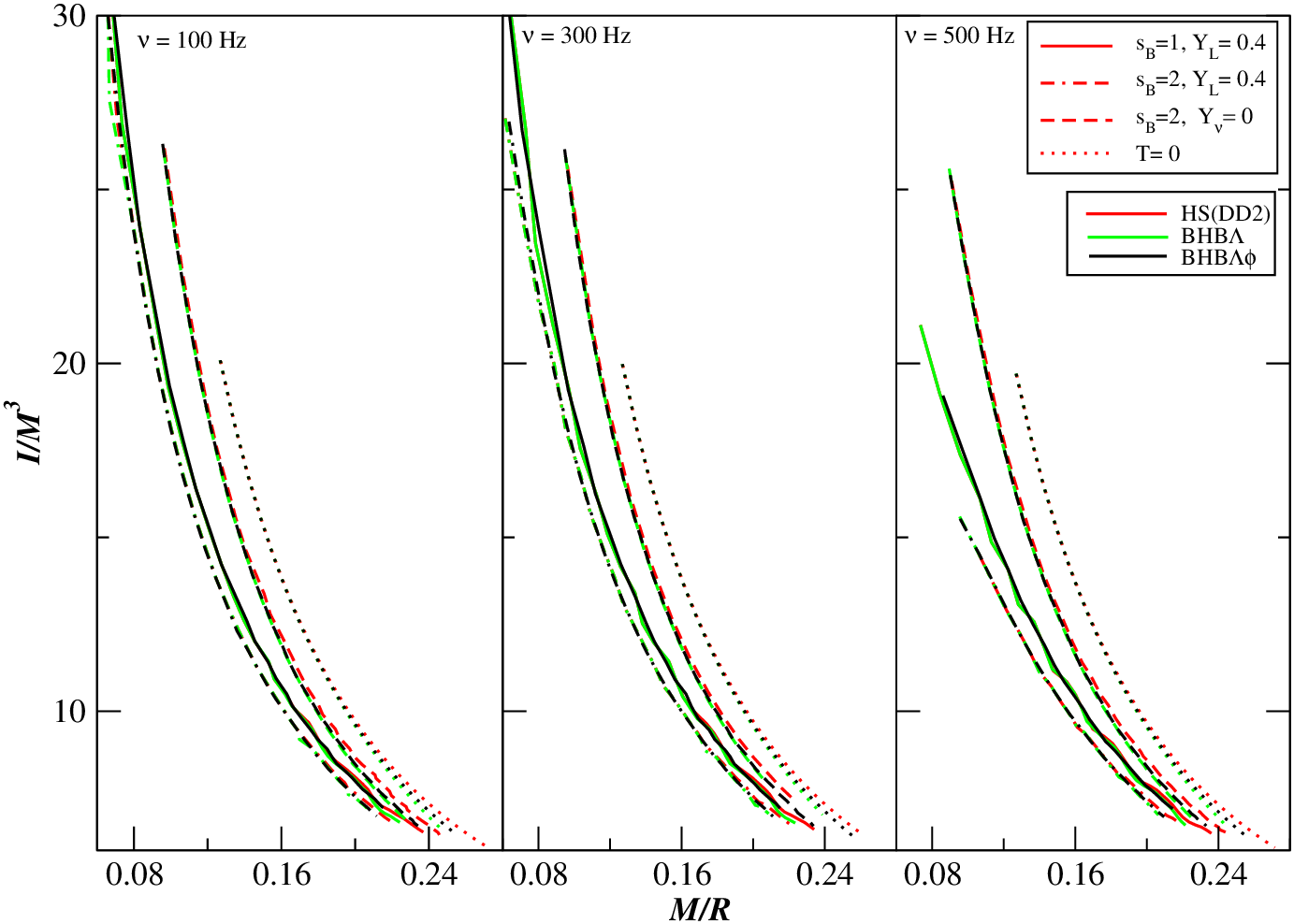}
\caption{Normalized moment of inertia ($I/M^3$) with compactness ($M/R$) for a star rotating at different frequencies as it evolves from PNS to NS.}
\label{Fig5}
\end{figure}

\begin{figure}
\includegraphics[width=\textwidth]{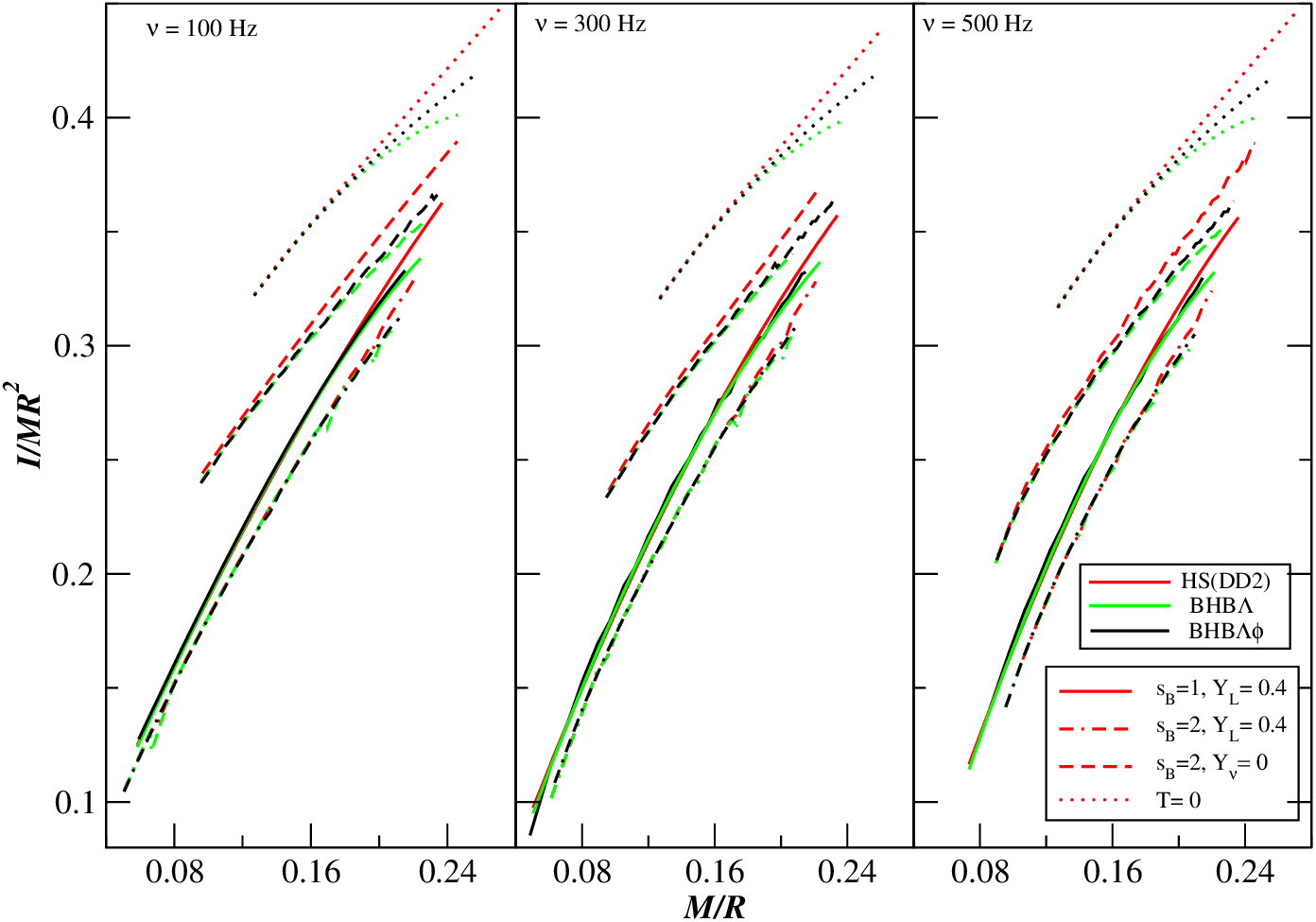}
\caption{Normalized moment of inertia ($I/MR^2$) variation with compactness ($M/R$) for same cases as in Fig.\ref{Fig5}.}
\label{Fig6}
\end{figure}

In Figs.~\ref{Fig5} and \ref{Fig6} we explore the universality relations for normalized moment of inertia ($\bar{I}$) with HS(DD2), BHB$\Lambda$ and BHB$\Lambda \phi$ EOSs corresponding to the four stages of PNS with respect to compactness $\mathcal{C}$. $\boldmath I$ is normalized to $M^3$ and $MR^2$ respectively in the two figures. Both the figures have three panels indicating three different spin frequencies from left to right i.e. 100, 300 and 500 Hz. We find the $\bar{I}$ lines are almost independent of the composition of the star corresponding to each PNS stage. But the lines corresponding to different temperature and lepton fraction are distinctly separated. This pattern is seen for both types of normalization and also for a particular frequency. Therefore, we might attribute this behavior to the combined effects of temperature and lepton fraction. 

\begin{figure}
\includegraphics[width=\textwidth]{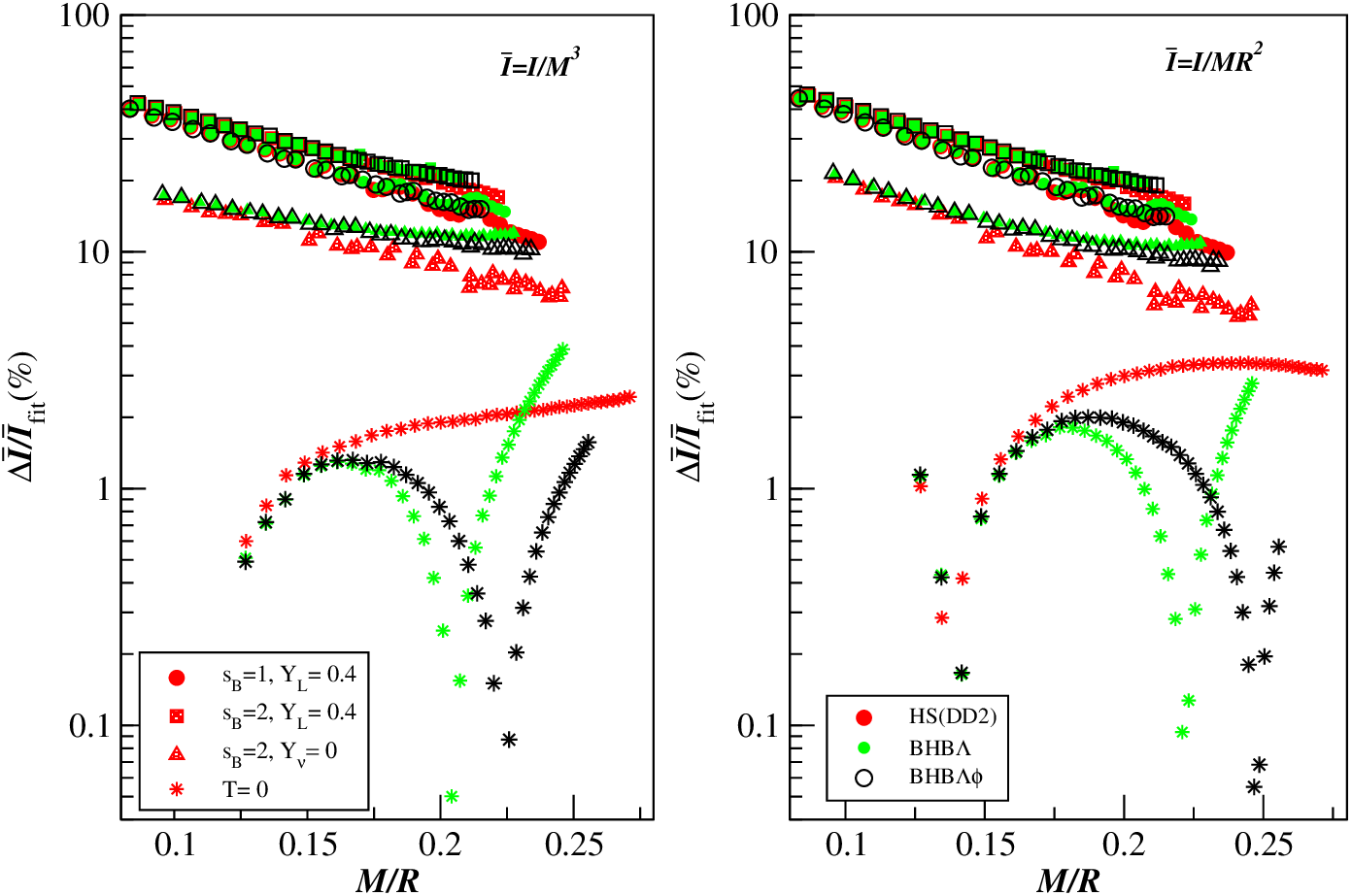}
\caption{Relative differences $\Delta\bar{\it I}/\bar {\it I_{fit}}$ = $|\bar{\it I}-\bar{\it I_{fit}}|/\bar{\it I_{fit}}$ for a slowly rotating 
PNS at different evolutionary stages; in the left panel $\bar{\it I}=\it I/M^3$ and in the right panel $\bar{\it I} =\it I/MR^2$. Different color schemes and different symbols are used for different EOSs and evolutionary stages respectively. }
\label{Fig7}
\end{figure}

Next, we try to quantify the deviations by comparing our calculated values of $\bar{I}$, plotted in Fig.~\ref{Fig5} and Fig.~\ref{Fig6}, for a 
star rotating at a fixed frequency of 100Hz with the ones we get from the fitting functions given by Breu and Rezzolla \cite{breu2016}. The fitting relations 
for $\bar{I}$ are,
\begin{equation}
\bar{I}_{fit}:=\frac{I}{MR^2}= a_0 + a_1 \mathcal{C} + a_4 \mathcal{C}^4
\end{equation}
and
\begin{equation}
\bar{I}_{fit}:=\frac{I}{M^3}= \bar{a}_1 \mathcal{C}^{-1} +  \bar{a}_2 \mathcal{C}^{-2} +  \bar{a}_3 \mathcal{C}^{-3} + \bar{a}_4 \mathcal{C}^{-4}.
\end{equation}
The values of the constants of Eqn. 8 are $a_0=0.244$, $a_1=0.638$, and $a_4=3.202$  \cite{breu2016}. 
The corresponding values for Eqn 9 are $\bar{a}_1=8.134\times 10^{−1}$, $\bar{a}_2=2.101\times 10^{−1}$, $\bar{a}_3=3.175\times 10^{−3}$, and $\bar{a}_4=-2.717\times 10^{-4}$  \cite{breu2016}. We use our calculated $\bar{I}$ data and compare those with the values (${\bar{I}}_{fit}$) obtained from Eqns. 8 and 9. We plot their relative differences $\Delta\bar{\it I}/{\bar {\it I}}_{fit}$ = $|\bar{\it I}-{\bar{\it I}}_{fit}|/{\bar{\it I}}_{fit}$ in Fig. \ref{Fig7}.  We find very high deviations ($\sim 40-50\%$), particularly in case of  high entropy and lepton rich EOS for both the normalization. This deviation is more evident for less compact stars. The deviation becomes smaller i.e. around $\sim 10\%$ for the case of $s_B=2$, $Y_\nu=0$. However, the deviations  are always the least, i.e. below $\sim 2-3\%$, for each EOS at T=0. Similar trends have been noticed for the stars rotating at 300 and 500 Hz. 

\begin{figure}
\includegraphics[width=\textwidth]{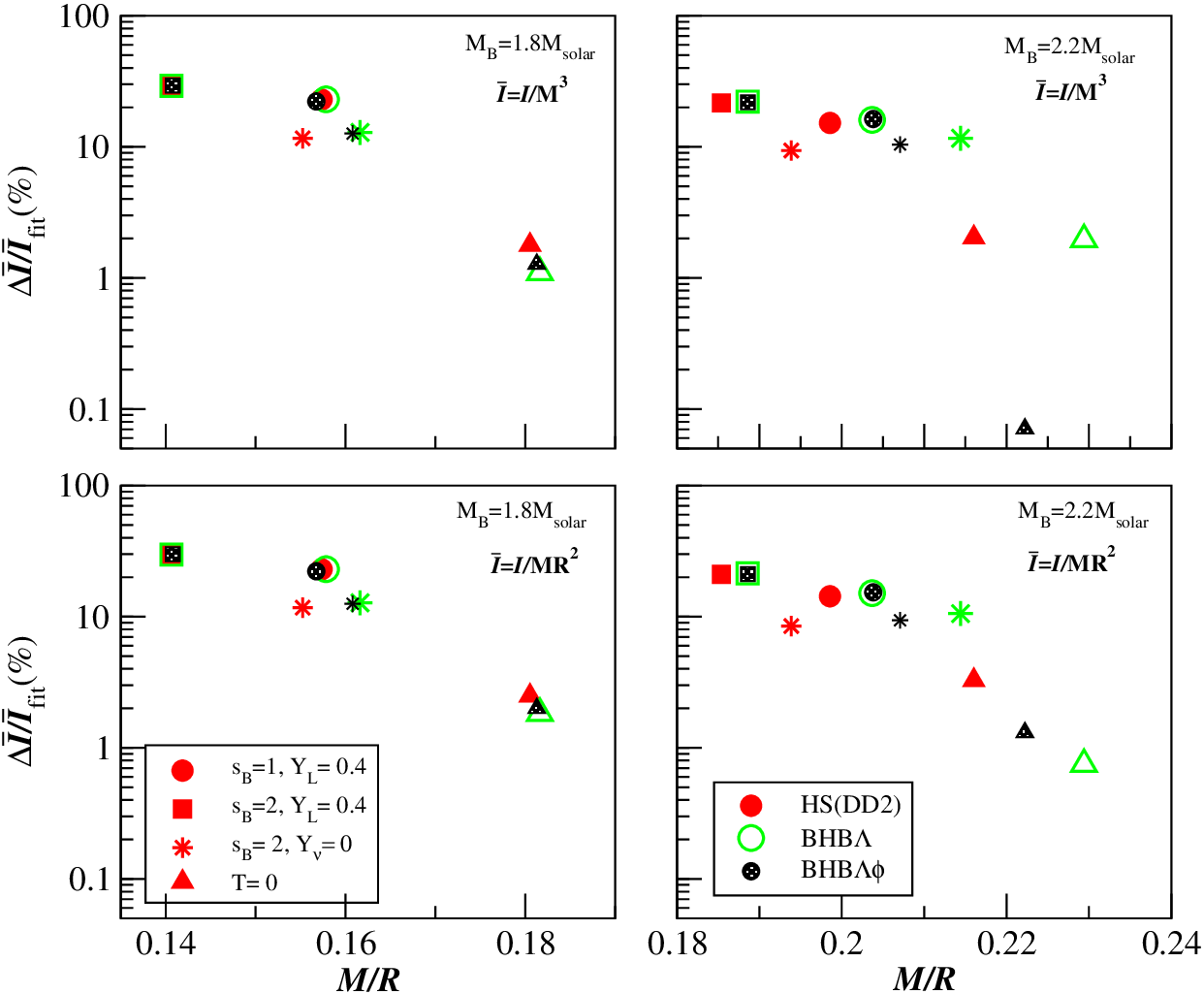}
\caption{ Relative differences $\Delta\bar{\it I}/\bar {\it I_{fit}}$ = $|\bar{\it I}-\bar{\it I_{fit}}|/\bar{\it I_{fit}}$ for a slowly rotating PNS at different evolutionary stages; in the upper panels 
$\bar{\it I}=\it I/M^3$, whereas in the lower panel $\bar{\it I} =\it I/MR^2$. Different color schemes and different symbols are used for different EOSs and evolutionary stages respectively. The two columns are for
fixed baryon mass $M_B= 1.8 M_{solar}$ and $2.2 M_{solar}$. }
\label{Fig8}
\end{figure}

Next, we consider the stages of a rotating PNS at fixed $M_B$ $1.8$ M$_{solar}$ and $2.2$ M$_{solar}$. At each stage, we measure the deviations from universality as done by Martinon et al. in the context of $I$-Love-$Q$ relations \cite{Marti2014}. Again we use the aforementioned fitting functions and fitting factors for $\bar I$. We plot these results in Fig.~\ref{Fig8} for a star rotating at a very low frequency of 5 Hz. As evident from Tables \ref {tab3} and \ref{tab4}, this is less than 1\% of the average Kepler limits. Here, we have used different symbols to distinguish the four evolutionary stages and different color schemes for the three chosen EOSs. For the $M_B=1.8$ $M_{solar}$ star, we don't find the deviations to be sensitive to the composition of the star except for the cold catalyzed one. The values of relative differences start with $\sim 22\%$ for stage I, followed by $\sim 30\%$ and $\sim 12\%$ for stage II and III respectively. Finally, when it reaches the cold catalyzed stage, the deviation falls down to $\sim 2\%$. Thus we conclude the $\bar{I}-\mathcal{C}$ relation is broken at the early stages of the life of a PNS, but the universality is restored once the star  becomes cold and attains beta stability. Our result is consistent with the earlier findings of Martinon et al. regarding the $\bar{I}-\bar{Q}$ relation \cite{Marti2014}. They have used the entropy and lepton fraction profiles from simulation results of evolving PNS at different time steps. On the other hand, Marques et. al \cite{Marq17} have  used arbitrary profile for entropy  but not considered any neutrino-trapped EOS in their calculations.
We take constant entropy per baryon throughout the star. Our lepton-rich matter at stages I and II contains trapped neutrinos. The deviations in the universality can be due to the combined effect of neutrino and thermal pressure. For the $M_B=2.2$ $M_{solar}$ star, the situation is almost similar. The only difference is that there is a deviation for different EOSs in the cold NS stage also. Nevertheless, they remain below $\sim 2\%$. Thus the conclusion remains the same.


\section{Summary}\label{sum}

In the present work,  we use the ad hoc profile which qualitatively represents the different evolutionary stages \cite{PrakashPR97, Burr86} and study the properties of a massive PNS containing $\Lambda$ hyperons. This is done using EOS within the framework of RMF model with density dependent couplings. The model uses the parameters which are consistent with several nuclear physics experimental data and astrophysical observational data. We construct the EOSs for $s_B=1 ~\textnormal{and} ~Y_L=0.4$; $s_B=2 ~\textnormal{and} ~Y_L=0.4$; $s_B=2 ~\textnormal{and} ~Y_\nu = 0$; cold-catalyzed  for both nucleonic and hyperonic models. The initial stages I and II contain electron-type neutrinos trapped in the PNS core. We calculate the M-R sequence for static star with those EOSs. We find a clear effect of temperature on the size of the stars. The hotter the star, the larger is the radius. The effect of $\phi$ mesons is evident on the temperature as well as on M-R relations. The difference between hyperonic and nucleon-only stars is quite visible in this model. We also note that the properties of less compact stars are governed mostly by their temperature and lepton content. The effect of neutrinos is evident in the EoS, structural properties and composition of the compact star matter. Several global properties of PNS are studied using those EOSs from static to maximally rotating configurations for fixed baryon masses of $1.8 ~\textnormal{and}~ 2.2 ~M_{solar}$. We see qualitative similarities for both the cases.

Another important finding of our studies is the deviation from $\bar{I}-{\cal C}$ universal relations for very hot and neutrino-rich stars. The temperature dependence in the $\bar{I}-{\cal C}$ relations was carried out here for the first time.  We take a slowly rotating star and measure the deviations for each of the stages.  We have seen with the arbitrary profile, the deviation is maximal in the first  three stages, while  universality is satisfied at cold catalysed stage.
Therefore, applying any universal relation to make a connection between a quantity measured before merger (e.g. tidal deformability $\kappa_2^T$) and another quantity after merger (e.g. peak frequency $f_2$) requires utmost caution. 

\section*{Acknowledgement}
The authors would like to thank D. Bandyopadhyay and R. Nandi for useful comments on the manuscript. S. S. Lenka would like to acknowledge the support from DST, India through INSPIRE fellowship.

\newpage

\vskip 2cm
\begin{table}
\caption{\label{tab3}Properties of non-rotating and 
rotating PNSs at the limiting frequency,
for a fixed baryonic mass \mbox{$M_B=1.8$ $ M_{solar}$}.
The parameters in the table are: 
central baryon number density ($n_{\rm B}$) [fm$^{-3}$],
gravitational mass ($M$) [$M_{solar}$], 
circumferential equatorial radius ($R_{\rm eq}$)[km],
Kepler frequency ($\nu_{\rm K}$)[Hz],
angular momentum ($J$)[$M_{solar}^2$], polar to equatorial axis ratio ($r_p$/$r_{eq}$)
and the rotation parameter ($|{\cal T}/W|$).}
\par
\begin{tabular}{c | c | c c c | c c c c c c c }
  \hline
  \hline
 EOS & PNS  & & $\nu = 0$ & & & & & $\nu$ = $\nu_{\rm K}$ & &  \\
     \cline{3-12}
   &  \small {stages}  & $n_{\rm B}$ & $M$ & $R_{\rm eq}$ & 
          $n_{\rm B}$ & $M$ & $R_{\rm eq}$ & 
        $\nu_{\rm K}$ & $J$ & $r_p$/$r_{eq}$ & $|{\cal T}/W|$ \\
\hline
\small {HS(DD2)}  & I   & 0.388 & 1.690 &  15.857 & 0.353 & 1.703 & 22.529  & 703 & 1.388 & 0.612 & 0.054 \\
 & II   & 0.346 & 1.711 &  17.969 & 0.322 & 1.723 & 25.909 & 577 & 1.244 & 0.623 & 0.042 \\
 & III   & 0.337 & 1.668 &  15.864 & 0.287 & 1.687 & 22.721 & 696  & 1.674 & 0.598 & 0.076 \\
 & IV   & 0.386 & 1.619 &  13.246 & 0.331 & 1.652 & 18.551 & 935  & 1.957 & 0.558 & 0.115 \\
\hline
\small {BHB$\Lambda$} & I    &  0.404  &  1.690  &  15.807   &  0.353  &  1.703  &  22.518   &   703 &  1.390  &  0.613  &  0.054  \\
 & II    &  0.346  &  1.711  &   17.970  &  0.322  &  1.724  &  25.909  &   577 &  1.245  &  0.623  &  0.042  \\
 & III    &  0.409  &  1.664  &   15.205  &  0.333  &  1.682  &  22.042  &  726   &  1.629  &  0.599  &  0.074  \\
 & IV   &  0.436  &  1.619  &   13.165  &  0.332  &  1.652  &  18.636  &  935   &  1.957  &  0.555  &  0.115  \\
\hline
\small {BHB$\Lambda$$\phi$} & I     &  0.394  &  1.687  &   15.894  &  0.354  &  1.699  &  22.553  &  702   &  1.357  &  0.615  &  0.054  \\
& II     &  0.347  &  1.713  &   17.964  &  0.322  &  1.723  &  25.893  &  576   &  1.242 &  0.625  &  0.042  \\
& III     &  0.386  &  1.664  &   15.283  &  0.326  &  1.686  &  22.167  &  724  &  1.633  &  0.597  &  0.074  \\
& IV    &  0.419  &  1.619  &   13.191  &  0.332  &  1.652  &  18.656  &  935   &  1.958  &  0.555  &  0.115  \\

    \hline
  \end{tabular}
\end{table}
\begin{table}
\caption{\label{tab4} 
Same as Table \ref{tab3}, but for  \mbox{$M_B= 2.2 M_{solar}$}.  }
\par
\begin{tabular}{c | c | c c c | c c c c c c c }
  \hline
  \hline
 EOS & PNS  & & $\nu = 0$ & & & & & $\nu$ = $\nu_{\rm K}$ & &  \\
     \cline{3-12}
   &  stages  & $n_{\rm B}$ & $M$ & $R_{\rm eq}$ & 
          $n_{\rm B}$ & $M$ & $R_{\rm eq}$ & 
        $\nu_{\rm K}$ & $J$ & $r_p$/$r_{eq}$ & $|{\cal T}/W|$ \\
\hline
HS(DD2)  & I   & 0.464 & 2.002 &  14.90 & 0.410 & 2.034 & 20.869  & 851 & 2.094 & 0.610 & 0.066 \\
 & II   & 0.436 & 2.04 &  16.244 & 0.389 & 2.065 & 23.184 & 735 & 1.958 & 0.616 & 0.055 \\
 & III   & 0.405 & 1.993 &  15.177 & 0.339 & 2.021 & 21.428 & 817  & 2.433 & 0.594 & 0.085 \\
 & IV   & 0.453 & 1.930 & 13.192 & 0.371 & 1.982 &  18.491 & 1028 & 2.815 & 0.550 & 0.124   \\
\hline
BHB$\Lambda$ & I    &  0.558  &  1.993  &  14.451   &  0.447  &  2.035 &  20.784   &   862 &  2.102 &  0.603 &  0.067  \\
 & II    &  0.522  &  2.042  &   15.987  &  0.389  &  2.064  &  23.159  &   735 &  1.955  &  0.617  &  0.054  \\
 & III    &  0.669  &  1.982  &   13.649  &  0.437  &  2.02  &  20.409  &  885   &  2.373  &  0.592  &  0.084  \\
 & IV   &  0.726  &  1.924  &   12.387  &  0.404  &  1.983  &  18.409  &  1037   &  2.82  &  0.551  &  0.124  \\
\hline
BHB$\Lambda$$\phi$ & I     &  0.543  &  2.001  &   14.494  &  0.442  &  2.036  &  20.722  &  860   &  2.087  &  0.608  &  0.066  \\
& II     &  0.501  &  2.042  &   16.035  &  0.388  &  2.064  &  23.331  &  735   &  1.96  &  0.611  &  0.055  \\
& III     &  0.545  &  1.981 &   14.128  &  0.424  &  2.018  &  20.554  &  874  &  2.389  &  0.593  &  0.085  \\
& IV    &  0.573  &  1.929  &   12.815  &  0.394  &  1.982  &  18.483  &  1034   &  2.82  &  0.549  &  0.124  \\

\hline
  \end{tabular}
\end{table}
\newpage


\end{document}